\documentclass[12pt]{article}
\usepackage{amssym}
\usepackage{graphics}

\def\bp{b^{\dagger}}
\def\C{\mbox{$\Bbb C$}}
\def\ap{a^{\dagger}}

\title{
Maths-type $q$-deformed coherent states for $q > 1$}

\author{C.\ Quesne $^{a,}$\thanks{Corresponding author.
{\sl E-mail addresses}: penson@lptl.jussieu.fr (K.A.\ Penson), cquesne@ulb.ac.be (C.\
Quesne), tkachuk@ktf.franko.lviv.ua (V.M.\ Tkachuk)}\ , K.A.\ Penson $^b$,  V.M.\ 
Tkachuk $^c$\\
{\small\sl $^a$ Physique Nucl\'eaire Th\'eorique et Physique Math\'ematique, 
Universit\'e Libre de Bruxelles,} \\ 
{\small \sl Campus de la Plaine CP229, Boulevard~du Triomphe, B-1050
Brussels, Belgium}\\
{\small\sl $^b$ Universit\'e Pierre et Marie Curie, Laboratoire de Physique Th\'eorique des
Liquides, }\\
{\small\sl CNRS UMR 7600, Tour 16, 5i\`eme \'etage, 4, place Jussieu, F-75252 Paris
Cedex 05, France}\\ {\small\sl $^c$ Ivan Franko Lviv National University, Chair of
Theoretical Physics,}\\ {\small\sl 12, Drahomanov Street, Lviv UA-79005, Ukraine}}
\date{ }
\begin{document}
\baselineskip=22pt plus 1pt minus 1pt
\maketitle

\begin{abstract}
Maths-type $q$-deformed coherent states with $q > 1$ allow a resolution of unity in the
form of an ordinary integral. They are sub-Poissonian and squeezed. They may be
associated with a harmonic oscillator with minimal uncertainties in both position and
momentum and are intelligent coherent states for the corresponding deformed Heisenberg
algebra.
\end{abstract}

\vspace{0.5cm}


 



\newpage
%
%
\section{Introduction}

Generalized coherent states (GCS), associated with algebras other than the oscillator one,
have attracted a lot of attention in connection with problems in quantum optics, as well
as in other areas ranging from solid state physics to cosmology (for a recent review
see~\cite{dodonov}). To be acceptable from a mathematical viewpoint~\cite{klauder63},
they have to satisfy a minimum set of conditions: (i) normalizability (as any
vector of a Hilbert space), (ii) continuity in their label, and (iii) existence of a
resolution of unity with a positive-definite weight function (implying that the states form
an (over)complete set).\par
%
%
An important class of GCS is provided by the $q$-deformed coherent states (CS), related
to deformations of the canonical commutation relation or, equivalently, to deformed
boson operators (see e.g.~\cite{odzi}). Among the latter, those satisfying the relation
\begin{equation}
  b \bp - q \bp b = I,  \label{eq:maths} 
\end{equation}
where $\bp$ is the Hermitian conjugate of $b$ and $q$ is some real constant~\cite{arik,
kuryshkin, jannussis}, are often termed maths-type $q$-bosons~\cite{solomon} because
the `basic' numbers and special functions associated with them have been investigated in
the mathematical literature for over 150 years (see e.g.~\cite{gasper}).\par
%
%
Eigenstates of the $q$-deformed annihilation operator $b$ of Eq.~(\ref{eq:maths}),
corresponding to some complex eigenvalue $z$, were introduced in~\cite{arik, jannussis}.
Such $q$-deformed CS were shown~\cite{arik} to provide a unity resolution relation
written as a $q$-integral with a weight function expressed in terms of some
$q$-exponential. An alternative formulation~\cite{kar} was proposed  in terms of an
ordinary integral with the corresponding weight function known through the formal inverse
Fourier transform of some given function. Its explicit form however cannot be easily
extracted for generic $q$. Such works were restricted to the parameter range $0 < q <
1$.\par
%
%
Some other approaches to the $q$-deformed CS associated with the maths-type
$q$-bosons, instead of employing the complex plane \C\ as the base field, used its
$q$-deformation $\C_q$, characterized by the re-ordering  rule $z z^* = q z^*
z$~\cite{brze}. Both the eigenstates of the annihilation operator $b$~\cite{kowalski}
and the states resulting from the action of a unitary $q$-analogue of the Weyl
displacement operator on the vacuum state~\cite{mcdermott} were studied in such a
framework.\par
%
%
In the present paper, we will revisit the problem of the annihilation operator eigenstates in
the conventional approach using the complex plane \C\ and consider this time the
`forgotten' parameter range $q > 1$. We actually plane to demonstrate that such
$q$-deformed CS have simpler mathematical properties than the corresponding ones for
$0 < q < 1$ in the sense that they allow a resolution of unity in terms of an ordinary
integral with a simple positive-definite weight function. In addition, we will show that they
have some nonclassical properties relevant to quantum optics and that they may have
some interesting applications in the context of a harmonic oscillator with minimal
uncertainties in both position and momentum~\cite{kempf94, kempf95, cq03}, as
suggested by some recent considerations in string theory and quantum gravity.\par
%
%
\section{Mathematical properties}

Let ${\cal F}_q$ be the $q$-deformed Fock space associated with the $q$-boson
creation and annihilation operators considered in Eq.~(\ref{eq:maths}). It is spanned by
the set of all $n$-$q$-boson states
\begin{equation}
  |n\rangle_q = \frac{(\bp)^n}{\sqrt{[n]_q!}}\, |0\rangle_q, \qquad n=0, 1, 2, \ldots,
  \label{eq:basis}
\end{equation}
where $|0\rangle_q$ is the normalized vacuum state, i.e., $b |0\rangle_q = 0$ and ${}_q
\langle 0|0 \rangle_q = 1$, and the
$q$-factorial
\begin{equation}
  [n]_q!  \equiv  \left\{\begin{array}{ll}
        1 & {\rm if\ } n=0, \\[0.2cm]
        [n]_q [n-1]_q \ldots [1]_q & {\rm if\ } n=1, 2, \ldots,
     \end{array}\right.  \label{eq:q-fac}
\end{equation}
is defined in terms of the $q$-numbers
\begin{equation}
  [n]_q \equiv \frac{q^n - 1}{q-1} = 1 + q + q^2 + \cdots + q^{n-1}.
  \label{eq:number}
\end{equation}
Note that the $q$-factorial (\ref{eq:q-fac}) can also be written in terms of the
$q$-gamma function $\Gamma_{q^{-1}}(n+1)$~\cite{gasper} as
\begin{equation}
  [n]_q! = q^{n(n-1)/2} \Gamma_{q^{-1}}(n+1).  \label{eq:q-gamma} 
\end{equation}
The states $|n\rangle_q$ form an orthonormal basis of ${\cal F}_q$,
\begin{equation}
  {}_q\langle n'|n \rangle_q = \delta_{n',n},
\end{equation}
and the action of the creation and annihilation operators on them is given by
\begin{equation}
  \bp |n\rangle_q = \sqrt{[n+1]_q}\, |n+1\rangle_q, \qquad b |n\rangle_q =
  \sqrt{[n]_q}\, |n-1\rangle_q.  
\end{equation}
\par
%
%
As well known, the operators $\bp$, $b$ can be realized in terms of conventional boson
operators $\ap$, $a$, satisfying the relation $[a, \ap] = I$, as
\begin{equation}
  \bp = \sqrt{\frac{[N]_q}{N}}\, \ap, \qquad b = a \sqrt{\frac{[N]_q}{N}},
  \label{eq:b-a}
\end{equation}
where $N \equiv \ap a$ and $[N]_q$ is defined as in (\ref{eq:number}). As a
consequence of (\ref{eq:b-a}), $\bp b = [N]_q$ and $b \bp = [N+1]_q$. By using such
a map and assuming $|0\rangle_q = |0\rangle$ (where $a |0\rangle = 0$), we can
actually identify $|n\rangle_q$ with the $n$-boson state $|n\rangle = (n!)^{-1/2}
(\ap)^n |0\rangle$ and ${\cal F}_q$ with the standard Fock space $\cal F$.\par
%
%
In ${\cal F}_q$, we may now look for the eigenstates of $b$ corresponding to some
complex eigenvalue $z$,
\begin{equation}
  b |z\rangle_q = z |z\rangle_q.  \label{eq:CS-def}
\end{equation}
They are easily constructed in terms of the $n$-$q$-boson states (\ref{eq:basis}) and are
given by
\begin{equation}
  |z\rangle_q = {\cal N}_q^{-1/2}(|z|^2) \sum_{n=0}^{\infty} \frac{z^n}{\sqrt{[n]_q!}}\,
  |n\rangle_q,  \label{eq:CS}
\end{equation}
where
\begin{equation}
  {\cal N}_q(t) = \sum_{n=0}^{\infty} \frac{t^n}{[n]_q!} = E_q(t)  \label{eq:q-exp}
\end{equation}
is some deformation of the exponential function $e^t$, to which it reduces for $q \to
1$ (we should add that the notation $E_q(t)$ is also used for other types of
deformations~\cite{gasper}).\par
%
%
In the cases considered so far~\cite{arik, jannussis, kar}, the deformation parameter $q$
has been restricted to the range $0 < q < 1$. Since for such values, $[n]_q < n$ for
$n=2$, 3,~\ldots, the convergence of the series defining $E_q(t)$ is slower than that of
the exponential series. As a consequence, the states $|z\rangle_q$ are normalizable only
on the disc of radius $[\infty]_q^{1/2} = (1-q)^{-1/2}$.\par
%
%
In contrast, for $q$ values greater than 1, to which we restrict ourselves here, $[n]_q >
n$ for $n=2$, 3,~\ldots, so that the convergence of $E_q(t)$ is faster than that of the
exponential and the states $|z\rangle_q$ turn out to be normalizable on the whole
complex plane.\par
%
%
They are also continuous in their label $z$, i.e.,
\begin{equation}
  [z - z'| \to 0 \qquad \Rightarrow \qquad \Bigl| |z\rangle_q - |z'\rangle_q \Bigr|^2 \to 0,
\end{equation}
because
\begin{equation}
  \Bigl| |z\rangle_q - |z'\rangle_q \Bigr|^2 = 2 (1 - {\rm Re}\, {}_q \langle z'|z \rangle_q)
\end{equation}
and
\begin{equation}
  {}_q \langle z'|z \rangle_q = \left[E_q(|z|^2) E_q(|z'|^2)\right]^{-1/2}
  E_q(z^{\prime*}z)  \label{eq:overlap}   
\end{equation}
is a continuous function.\par
%
%
Moreover, they give rise to a resolution of unity in ${\cal F}_q$,
\begin{equation}
  \int\int_{\C} d^2z\, |z\rangle_q\, W_q(|z|^2)\, {}_q\langle z| = \sum_{n=0}^{\infty}
  |n\rangle_q\, {}_q\langle n| = I,  \label{eq:resolution}
\end{equation}
with a positive-definite weight function
\begin{equation}
  W_q(|z|^2) = \frac{q-1}{\pi \ln q} \frac{E_q(|z|^2)}{E_q(q|z|^2)}.  \label{eq:weight}
\end{equation}
\par
%
%
To prove (\ref{eq:resolution}) and (\ref{eq:weight}), we transform the former into the
Stieltjes power-moment problem
\begin{equation}
  \int_0^{\infty} dt\, t^n {\tilde W}_q(t) = [n]_q!, \qquad n=0, 1, 2, \ldots, 
  \label{eq:power}
\end{equation}
by using Eq.~(\ref{eq:CS}) and its Hermitian conjugate. In (\ref{eq:power}),
${\tilde W}_q(t)$ is defined by ${\tilde W}_q(t) = \pi W_q(t)/{\cal N}_q(t)$, where $t
\equiv |z|^2$. Equation~(\ref{eq:power}) is a special case of  
\begin{equation}
  \int_0^{\infty} dt\, t^{s-1} {\tilde W}_q(t) = q^{(s-1)(s-2)/2} \Gamma_{q^{-1}}(s),
  \qquad {\rm Re}\,s > 0,  \label{eq:power-bis}
\end{equation}
where Eq.~(\ref{eq:q-gamma}) has been taken into account.
Equation~(\ref{eq:power-bis}) as it stands is valid for all $q > 1$.\par
%
%
In order to obtain a solution $\tilde{W}_q(t)$ of Eq.~(\ref{eq:power-bis}), we return to
the $q < 1$ case by making a transformation $q' = 1/q$ ($q' < 1$), supplemented by a
change of variable $t' = t/q'$. This yields
\begin{equation}
  \int_0^{\infty} dt'\, t^{\prime s-1} {\tilde W}_{q^{\prime -1}}(q't') = q^{\prime-1}
  q^{\prime -s(s-1)/2} \Gamma_{q'}(s), \qquad {\rm Re}\,s > 0.
\end{equation}
This last form allows one to make a link to Eq.~(3.9) of~\cite{ataki}, which directly
furnishes
\begin{equation}
  \tilde{W}_{q^{\prime-1}}(q't') = \frac{1-q'}{q' \ln(q^{\prime-1})}
  \left\{E^J_{q'}[(1-q')t']\right\}^{-1},
\end{equation}
where $E^J_{q'}(x)$ is one of Jackson's $q'$-exponential (Eq.~(1.3.16)
of~\cite{gasper}). When rewritten with $q= 1/q' > 1$ again, this becomes
\begin{equation}
  \tilde{W}_q(t) = \frac{q-1}{\ln q} \left\{E^J_{q^{-1}}[(q-1)t)]\right\}^{-1} =
\frac{q-1}{\ln q} [E_q(qt)]^{-1},  \label{eq:tilde-W}
\end{equation}
where in the last step we used the relation between Jackson's $q$-exponential and that
defined in Eq.~(\ref{eq:q-exp}). The solution (\ref{eq:tilde-W}) of the power-moment
problem (\ref{eq:power}) is a manifestly positive function, which is illustrated for a few $q$
values in Fig.~1. Observe that for $q \to 1$, $\tilde{W}_q(t) \to \tilde{W}(t) = e^{-t}$. 
\par
%
%
We therefore conclude that the states $|z\rangle_q$, defined in (\ref{eq:CS}), qualify as
GCS in the sense described in Sec.~1. It is worth stressing, however, that contrary to the
conventional CS, but in analogy with many GCS (see e.g.~\cite{sixdeniers, cq01, cq02}), it
can be shown that the solution $\tilde{W}_q(t)$ of the power-moment
problem~(\ref{eq:power}) may not be unique. Methods similar to those used
in~\cite{sixdeniers} might be employed to construct other solutions.\par
%
%
Since the states $|z\rangle_q$ form a complete (in fact, an overcomplete) set in ${\cal
F}_q$ when $z$ runs over the complex plane, we can associate with them a realization of
${\cal F}_q$ as a space ${\cal B}_q$ of entire analytic functions, i.e., a $q$-deformed
Bargmann representation.\par
%
%
{}For such a purpose, it is convenient to replace the normalized CS $|z\rangle_q$ by
unnormalized ones (denoted by a round bracket instead of an angular one), $|z)_q \equiv
{\cal N}_q^{1/2}(|z|^2) |z\rangle_q$, and to substitute $\xi^*$ for $z$. Then any vector
$|\psi\rangle_q = \sum_{n=0}^{\infty} c_n(q) |n\rangle_q \in {\cal F}_q$, i.e., such that
$\sum_{n=0}^{\infty} |c_n(q)|^2 < \infty$, can be realized by the entire function
\begin{equation}
  \psi_q(\xi) = {}_q(\xi^*|\psi\rangle_q = \sum_{n=0}^{\infty}
  \frac{c_n(q)}{\sqrt{[n]_q!}} \xi^n.
\end{equation}
The functions $\psi_q(\xi)$ are the elements of a Hilbert space ${\cal B}_q$, whose scalar
product is defined by
\begin{equation}
  (\psi'_q, \psi_q) = \frac{1}{\pi} \int\int_{\C} d^2\xi\, {\tilde W}_q(|\xi|^2)
  [\psi'_q(\xi)]^* \psi_q(\xi). 
\end{equation}
\par
%
%
Any operator $O$, defined in ${\cal F}_q$, is represented in ${\cal B}_q$ by some
$q$-differential operator ${\cal O}_q$, defined by ${}_q(\xi^*| O |\psi\rangle_q = {\cal
O}_q \psi_q(\xi)$ for any $|\psi\rangle_q \in {\cal F}_q$. In particular, $\bp$ and $b$
are represented by the operator acting in ${\cal B}_q$ as a multiplication by $\xi$ and by
the $q$-differential operator ${\cal D}_q$, defined by ${\cal D}_q \psi_q(\xi) =
[\psi_q(\xi) - \psi_q(q\xi)]/[(1-q)\xi]$, respectively.\par
%
%
In the special case where $|\psi\rangle_q$ is some coherent state $|z\rangle_q$, we
obtain from (\ref{eq:overlap}) that the corresponding function is $\psi_z(\xi) =
[E_q(|z|^2)]^{-1/2} E_q(z \xi)$. Such a result can also be derived from the realization of
Eq.~(\ref{eq:CS-def}) in ${\cal B}_q$, which writes ${\cal D}_q \psi_z(\xi) = \xi
\psi_z(\xi)$.\par
%
%
\section{Geometrical and physical properties in quantum optics}

Since through the mapping of $|n\rangle_q$ on $|n\rangle$, the $q$-deformed CS
$|z\rangle_q$ are special linear combinations of $n$-boson states, it is worth analyzing
their nonclassical properties in quantum optics. The simple structure of $|z\rangle_q$
makes it easy to calculate expectation values of operators. In the following, we
extensively use the relation~\cite{sixdeniers}
\begin{eqnarray}
  {}_q \langle (\ap)^p a^r\rangle_q & = & (z^*)^p z^r S^{(p,r)}_q(t), \nonumber \\
  S^{(p,r)}_q(t) & = & \frac{1}{{\cal N}_q(t)} \sum_{n=0}^{\infty} \left(\frac{(n+p)!
         (n+r)!}{[n+p]_q! [n+r]_q!}\right)^{1/2} \frac{t^n}{n!} \qquad r, p = 0, 1, 2, \ldots,
\end{eqnarray}
where ${}_q\langle O \rangle_q \equiv {}_q\langle z| O |z \rangle_q$ and $t \equiv
|z|^2$.\par
%
%
To start with, let us consider the metric factor $\omega_q(t) = d \langle
N\rangle_q/dt$, determining the geometry of the two-dimensional surface generated by
the map from $z \in \C$ to $|z\rangle_q \in {\cal F}_q$ and described in
polar coordinates $r$, $\theta$ (i.e., $z = r e^{{\rm i}\theta}$) by the line element
$d\sigma_q^2 = \omega_q(r^2) (dr^2 + r^2 d\theta^2)$~\cite{klauder01}. It can be
shown that for small $t$ values, $\omega_q(t) \simeq 1 - 2(q-1) (q+1)^{-1} t + O(t^2)
< \omega(t) = 1$, where the latter corresponds to the flat geometry of the conventional
CS. Numerical calculations confirm this result for higher $t$ values: $\omega_q(t)$
indeed turns out to be a decreasing function of $t$, the departure from $\omega(t) = 1$
increasing with the $q$ value.\par
%
%
Information about the photon number distribution is provided by the Mandel
parameter~\cite{mandel}
\begin{equation}
  Q_q = \frac{(\Delta N)_q^2 - \langle N \rangle_q}{\langle N \rangle_q}, \qquad
  (\Delta N)_q^2 \equiv \langle N^2 \rangle_q - \langle N \rangle_q^2,   
\end{equation}
which vanishes for the Poisson distribution, is positive for a super-Poissonian distribution
(bunching effect) and is negative for a sub-Poissonian one (antibunching effect). For
small $t$ values, we find $Q_q(t) \simeq - (q-1)t/(q+1) + O(t^2)$. In Fig.~2, we plot
$Q_q(t)$ as a function of $t$ for some $q$ values. We notice that it is always negative
and that consequently the state $|z\rangle_q$ is sub-Poissonian. Moreover, the
departures from the Poisson distribution, characterizing conventional CS, increase with
$q$.\par
%
%
The variances $(\Delta X)_q^2$ and $(\Delta P)_q^2$ of the Hermitian quadrature
operators $X = (a + \ap)/\sqrt{2}$, $P = (a - \ap)/({\rm i}\sqrt{2})$, in any state
$|\psi\rangle_q \in {\cal F}_q$ satisfy the conventional uncertainty relation $(\Delta
X)_q^2 (\Delta P)_q^2 \ge 1/4$. The lower bound is attained by the vacuum state, for
which both variances are equal to 1/2. A state $|\psi\rangle_q$ is squeezed for the
quadrature $X$ (resp.\ $P$) if $(\Delta X)_q^2 < 1/2$ (resp.\ $(\Delta P)_q^2 <
1/2$)~\cite{walls}.\par
%
%
{}For the $q$-deformed CS $|z\rangle_q$, one easily shows that
\begin{equation}
  (\Delta X)_q^2 = 2({\rm Re}\, z)^2 \left\{S^{(2,0)}_q(t) - \left[S^{(1,0)}_q(t)
  \right]^2\right\} + t \left[S^{(1,1)}_q(t) - S^{(2,0)}_q(t)\right] + \frac{1}{2},
  \label{eq:variance}
\end{equation}
and that $(\Delta P)_q^2$ is given by a similar expression with ${\rm Re}\, z$ replaced
by ${\rm Im}\, z$. It is therefore enough to study $(\Delta X)_q^2$. It can be checked
numerically that the coefficient of $2 ({\rm Re}\, z)^2$ in (\ref{eq:variance}), behaving
as $- \left[1 - \sqrt{2/(1+q)}\right] + O(t)$ for small $t$ values, is actually negative over
a wide range of $t$ values. Hence, for a given $t$, the maximum squeezing in $X$
can be achieved when $z$ is real. Let us therefore set $z = \sqrt{t}$ and consider the ratio
$R_q(t) = 2 (\Delta X)_q^2$ of the variance $(\Delta X)_q^2$ in $|z\rangle_q$ to the
variance 1/2 in the vacuum state. For small $t$ values, we find $R_q(t) \simeq 1 -
2\left[1 - \sqrt{2(1+q)}\right]t + O(t^2)$, showing the presence of squeezing. This is
confirmed in Fig.~3, where $R_q(t)$ is plotted against $t$ for $z = \sqrt{t}$ and
several $q$ values.\par
%
%
{}Finally, let us consider the signal-to-quantum noise ratio $\sigma_q = \langle X
\rangle_q^2/(\Delta X)_q^2$, where
\begin{equation}
  \langle X \rangle_q^2 = 2 ({\rm Re}\, z)^2 \left[S^{(1,0)}_q(t)\right]^2.
\end{equation}
For a given $t$ value, such a ratio is maximum for real $z$. Setting $z = \sqrt{t}$ again,
we obtain for small $t$, $\sigma_q(t) \simeq 4t [1 + O(t^2)]$. This may be compared
with $4 \langle N \rangle_q \simeq 4t [1 - (q-1)t/(q+1) + O(t^2)]$ and $4 \langle N
\rangle_q (\langle N \rangle_q + 1) \simeq 4t [1 + 2t/(q+1) + O(t^2)]$, corresponding
to the values $4N_s$ and $4N_s (N_s+1)$ attained for the conventional CS and the
conventional squeezed state, respectively (here $N_s = \langle N \rangle$ denotes the
number of photons in the signal). Hence there is an improvement on the conventional CS
value, but it still remains rather far from that achieved by the conventional squeezed
state, giving rise to the upper bound to $\sigma_q$~\cite{yuen}. This can again be
confirmed numerically over an extended range of $t$ values.\par
%
%
\section{Application to a harmonic oscillator with minimal uncertainties in both position
and momentum}

Studies on small distances in string theory and quantum gravity suggest the existence of
a finite lower bound to the possible resolution of length $\Delta x_0$. On large scales
there is no notion of plane waves or momentum eigenvectors on generic curved spaces
and this leads to the suggestion that there could exist a lower bound to the possible
resolution of momentum $\Delta p_0$. It is a natural, though nontrivial, assumption that
minimal length and momentum should quantum theoretically be described as nonzero
minimal uncertainties in position and momentum measurements.\par
%
%
Such minimal uncertainties can be described in the framework of small corrections to the
canonical commutation relation~\cite{kempf94}
\begin{equation}
  [x, p] = {\rm i} \hbar (1 + \alpha x^2 + \beta p^2) = {\rm i} \hbar \left[1 + (q-1)
  \left(\frac{x^2}{4L^2} + \frac{p^2}{4K^2}\right)\right],  \label{eq:def-alg} 
\end{equation}
where $\alpha \ge 0$, $\beta \ge 0$, and the reparametrization $\alpha =
(q-1)/(4L^2)$, $\beta = (q-1)/(4K^2)$ is used. Here the constants $L$, $K$ carry
units of length and momentum and are related by $4KL = \hbar (1+q)$, while
\begin{equation}
  q = \frac{1 + \hbar \sqrt{\alpha \beta}}{1 - \hbar \sqrt{\alpha \beta}}  \label{eq:q}
\end{equation}
is a dimensionless constant such that $q \ge 1$, which immediately brings us back to the
case considered in the previous sections. The minimal uncertainties in the position and the
momentum are then given by
$\Delta x_0 = L
\sqrt{(q-1)/q}$ and
$\Delta p_0 = K \sqrt{(q-1)/q}$, respectively.\par
%
%
The special case where $\alpha = 0$ and $\beta > 0$ can be treated by setting $K(q) =
\frac{1}{2} \sqrt{(q-1)/\beta}$, $L(q) = \frac{1}{2} \hbar (1+q) \sqrt{\beta/(q-1)}$,
and letting $q$ go to 1. The minimal uncertainty in the position then turns out to be
given by $\Delta x_0 = \hbar \sqrt{\beta}$, whereas there is no nonvanishing minimal
momentum uncertainty. As a consequence, the deformed Heisenberg algebra
(\ref{eq:def-alg}) can be represented on momentum space wave functions (although not
on position ones). Similarly, in the case where $\alpha > 0$ and $\beta = 0$, there is
only a nonvanishing minimal uncertainty in the momentum $\Delta p_0 = \hbar
\sqrt{\alpha}$ and the algebra (\ref{eq:def-alg}) can be represented on position space
wave functions.\par
%
%
In the general case where $\alpha > 0$ and $\beta > 0$, there is neither position nor
momentum representation so that one has to resort to a generalized Fock space
representation~\cite{kempf94}, wherein $x$ and $p$ are represented as
\begin{equation}
  x = L (\bp + b), \qquad p = {\rm i} K (\bp - b),  \label{eq:x-p}
\end{equation}
in terms of some operators $b$ and $\bp$ obeying Eq.~(\ref{eq:maths}) with $q$ given
by Eq.~(\ref{eq:q}).\par
%
%
Recently, there has been much interest in the harmonic oscillator
\begin{equation}
  H = \frac{p^2}{2m} + \frac{1}{2} m \omega^2 x^2,  \label{eq:h-o}
\end{equation}
where the position and momentum operators satisfy the deformed commutation
relation~(\ref{eq:def-alg}). The eigenvalue problem for such an oscillator was solved
exactly in the momentum representation in the special case where $\alpha = 0$ and
$\beta > 0$~\cite{kempf95}. Its solution in the general case where $\alpha > 0$ and
$\beta > 0$ will be given elsewhere~\cite{cq03}.\par
%
%
Here we would like to point out that on inserting (\ref{eq:x-p}) into (\ref{eq:h-o}) and
using (\ref{eq:maths}), $H$ can be transformed into
\begin{equation}
  H = \left(- \frac{K^2}{2m} + \frac{1}{2} m \omega^2 L^2\right) \left[(b^{\dagger})^2
  + b^2\right] + \left(\frac{K^2}{2m} + \frac{1}{2} m \omega^2 L^2\right)
  \left(b^{\dagger} b + b b^{\dagger}\right),
\end{equation}
which becomes
\begin{equation}
  H = \frac{1}{4} (1+q) \hbar\omega \{b, \bp\},  \label{eq:h-o-bis}
\end{equation}
provided the condition
\begin{equation}
  K = m\omega L \qquad {\rm or} \qquad \alpha = m^2 \omega^2 \beta
\end{equation}
is fulfilled. Such a condition can be achieved in two different ways: either by assuming a
specific relation between the deformation parameters $\alpha$, $\beta$ of
Eq.~(\ref{eq:def-alg}), but leaving the oscillator frequency $\omega$ arbitrary, or by
considering general deformation parameters $\alpha$, $\beta$, and an oscillator of
frequency $\omega_0 = \frac{1}{m} \sqrt{\frac{\alpha}{\beta}}$.\par
%
%
The Hamiltonian given in Eq.~(\ref{eq:h-o-bis}) is the most commonly used assumption
for the $q$-deformed harmonic oscillator Hamiltonian of frequency $\frac{1}{2} (1+q)
\omega$, associated with maths-type $q$-bosons. Its eigenstates are the
$n$-$q$-boson states (\ref{eq:basis}) and the corresponding eigenvalues are
\begin{equation}
  E_n(q) = \frac{1}{4}(1+q) ([n]_q + [n+1]_q) \hbar \omega = \frac{1}{4}(1+q)
  \{(1+q)[n]_q + 1\} \hbar \omega. 
\end{equation}
\par
%
%
The GCS $|z\rangle_q$, considered in the previous sections, are just the usual CS that
can be associated with a Hamiltonian of type (\ref{eq:h-o-bis}). From Eq.~(\ref{eq:x-p}), it
can be shown that the averages and variances of $x$ and $p$ in such states are given by
\begin{eqnarray}
  \langle x \rangle_q & = & 2L\, {\rm Re}\, z, \qquad \langle p \rangle_q = 2K\, {\rm
         Im}\, z, \nonumber \\
  (\Delta x)^2_q & = & L^2 [1 + (q-1)|z|^2], \qquad (\Delta p)^2_q = K^2 [1 +
         (q-1)|z|^2],   
\end{eqnarray}
respectively. This implies that the generalized uncertainty relation~\cite{kempf94}
\begin{equation}
  \Delta x \, \Delta p \ge \frac{\hbar}{2} \left\{1 + \alpha \left[(\Delta x)^2 + 
  \langle x \rangle^2\right] + \beta \left[(\Delta p)^2 +  \langle p \rangle^2\right]
  \right\},
\end{equation}
corresponding to the deformed commutation relation (\ref{eq:def-alg}), becomes an
equality or, in other words, that the states $|z\rangle_q$ are intelligent
CS~\cite{aragone} for the deformed Heisenberg algebra (\ref{eq:def-alg}).\par
%
%
We conclude that the states $|z\rangle_q$ may have some interesting applications in
the context of theories describing minimal uncertainties in both position and momentum,
as suggested by Eq.~(\ref{eq:def-alg}).\par
%
%
\section{Conclusion}

In the present paper, we have revisited the maths-type $q$-deformed CS with a special
emphasis on the `forgotten' case $q > 1$. We have proved that they are both
normalizable on the whole complex plane and continuous in their label, and that they
satisfy a unity resolution relation in the form of an ordinary integral with a
positive-definite weight function. Thus they give rise to a $q$-deformed Bargmann
representation of the maths-type $q$-boson operators.\par
%
%
In addition, we have investigated some characteristics of those CS relevant to quantum
optics. We have shown that they exhibit some nonclassical properties, such as
antibunching, quadrature squeezing, and enhancement of the signal-to-quantum noise
ratio. We have also pointed out that they induce a deformation of the metric.\par
%
%
{}Furthermore, we have established that they may be associated with a harmonic
oscillator with nonzero minimal uncertainties in both position and momentum, as resulting
from a deformed Heisenberg algebra in current use. We have finally shown that they are
intelligent CS for such an algebra.\par
%
%
\section*{Acknowledgments}

This work was initiated during a visit of C.Q.\ to the Universit\'e Pierre et Marie Curie,
Paris. C.Q.\ is a Research Director of the National Fund for Scientific Research (FNRS),
Belgium.\par
%
%

%
%
\newpage
\section*{Figure captions}

{}Fig.\ 1. The weight function $\tilde{W}_q(t)$ as a function of $t = |z|^2$ for $q = 1$
(solid line), $q = 1.5$ (dashed line), $q = 2$ (dotted line) and $q = 2.5$ (dot-dashed
line).

\noindent
{}Fig.\ 2. The Mandel parameter $Q_q(t)$ as a function of $t = |z|^2$ for $q = 1.1$
(solid line), $q = 1.2$ (dashed line) and $q = 1.3$ (dot-dashed line).

\noindent
{}Fig.\ 3. The variance ratio $R_q(t)$ as a function of $t = |z|^2$ for real $z$
and $q = 1.1$ (solid line), $q = 1.2$ (dashed line) and $q = 1.3$ (dot-dashed line).
\par
%
%
%
%
%
%
%
%
\newpage
\begin{picture}(160,100)
\put(35,0){\mbox{\scalebox{1.0}{\includegraphics{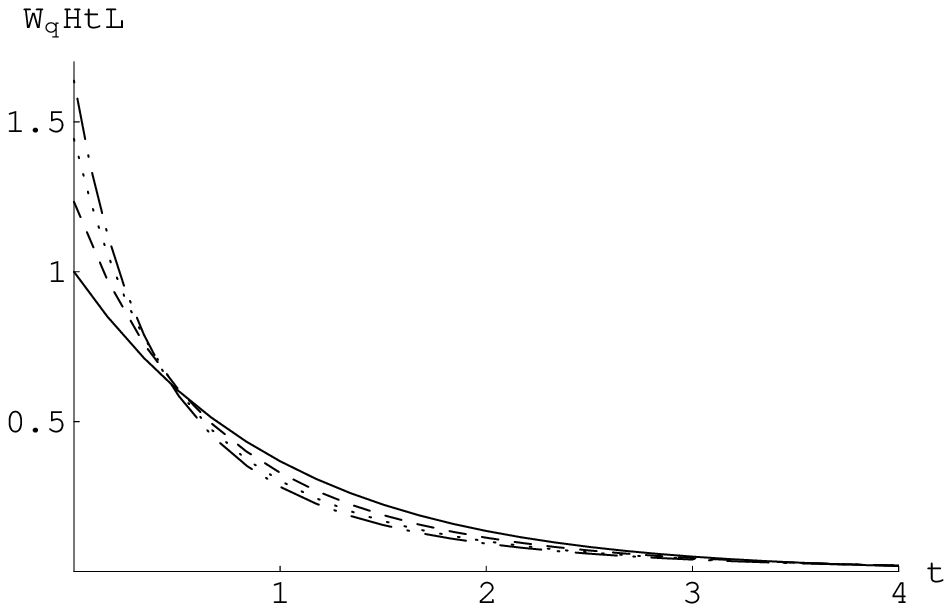}}}}
\end{picture}
%
%
\newpage
\begin{picture}(160,100)
\put(35,0){\mbox{\scalebox{1.0}{\includegraphics{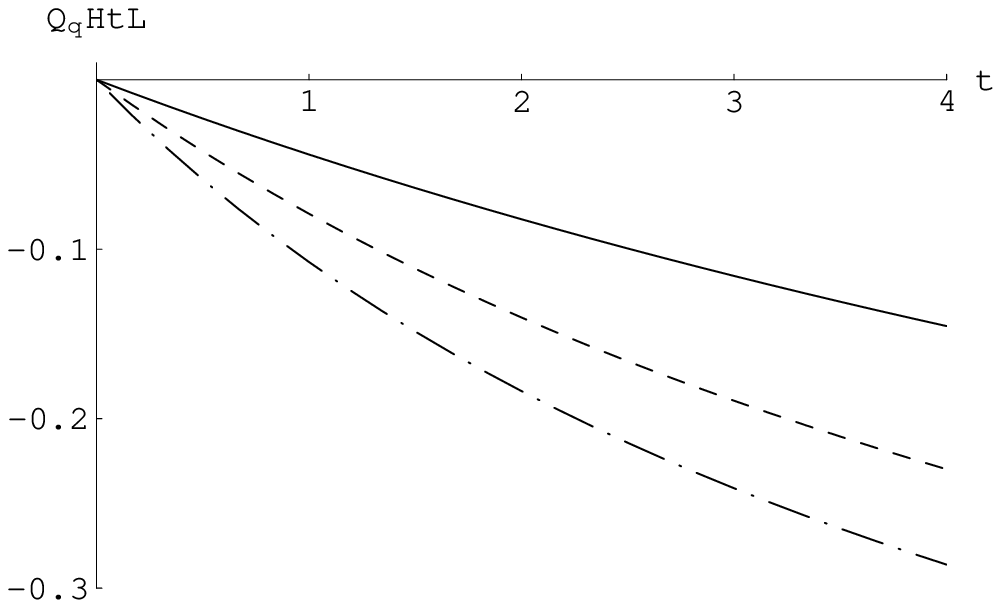}}}}
\end{picture}
%
%
\newpage
\begin{picture}(160,100)
\put(35,0){\mbox{\scalebox{1.0}{\includegraphics{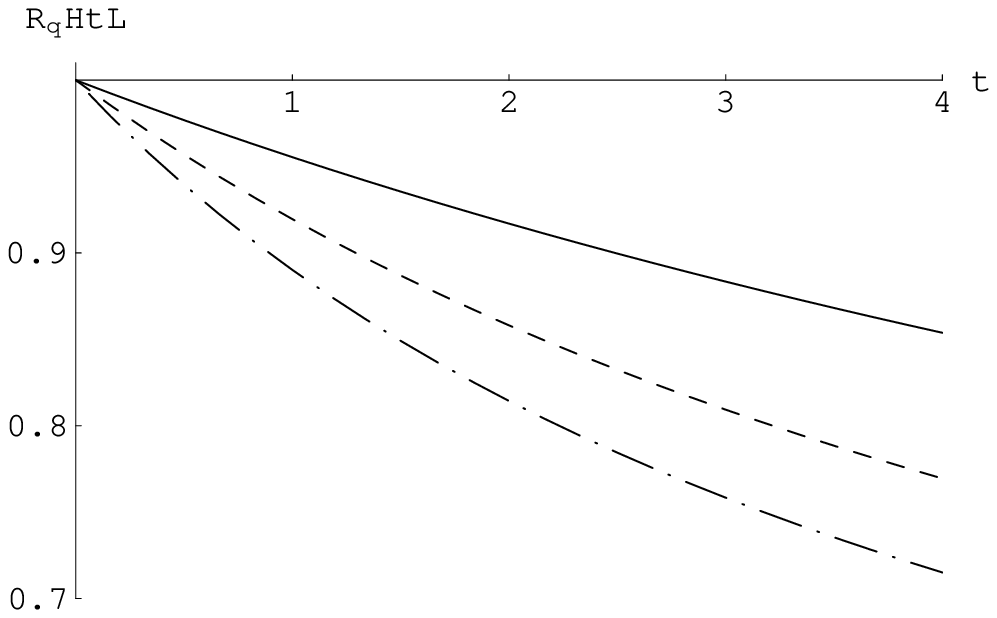}}}}
\end{picture}
%
%

\newpage
\begin{thebibliography}{99}

\bibitem{dodonov} V.V.\ Dodonov, Quantum Semiclass.\ Opt.\ 4 (2002) R1.

\bibitem{klauder63} J.R.\ Klauder, J.\ Math.\ Phys.\ 4 (1963) 1055, 1058.

\bibitem{odzi} A.\ Odzijewicz, Commun.\ Math.\ Phys.\ 192 (1998) 183.

\bibitem{arik} M.\ Arik and D.D.\ Coon, J.\ Math.\ Phys.\ 17 (1976) 524.

\bibitem{kuryshkin} V.\ Kuryshkin, Ann.\ Fond.\ L.\ de Broglie 5 (1980) 111.

\bibitem{jannussis} A.\ Jannussis, G.\ Brodimas, D.\ Sourlas and V.\ Zisis, Lett.\ Nuovo
Cimento 30 (1981) 123.

\bibitem{solomon} A.I.\ Solomon, Phys.\ Lett.\ A 196 (1994) 29.

\bibitem{gasper} G.\ Gasper and M.\ Rahman, Basic Hypergeometric Series, Cambridge
University Press, Cambridge, 1990.

\bibitem{kar} T.K.\ Kar and G.\ Ghosh, J.\ Phys.\ A 29 (1996) 125.

\bibitem{brze} T.\ Brzezi\'nski, H.\ D\c abrowski and J.\ Rembieli\'nski, J.\ Math.\
Phys.\ 33 (1992) 19;\\ 
T.\ Brzezi\'nski and J.\ Rembieli\'nski, J.\ Phys.\ A 25 (1992) 1945.

\bibitem{kowalski} K.\ Kowalski and J.\ Rembieli\'nski, J.\ Math.\ Phys.\ 34 (1993)
2153.

\bibitem{mcdermott} R.J.\ McDermott and A.I.\ Solomon, J.\ Phys.\ A 27 (1994) 2037.

\bibitem{kempf94}A.\ Kempf, J.\ Math.\ Phys.\ 35 (1994) 4483;\\
H.\ Hinrichsen and A.\ Kempf, J.\ Math.\ Phys.\ 37 (1996) 2121.

\bibitem{kempf95} A.\ Kempf, G.\ Mangano and R.B.\ Mann, Phys.\ Rev.\ D 52 (1995)
1108;\\
L.N.\ Chang, D.\ Minic, N.\ Okamura and T.\ Takeuchi, Phys.\ Rev.\ D 65 (2002) 125027.

\bibitem{cq03} C.\ Quesne and V.M.\ Tkachuk, in preparation.

\bibitem{ataki} N.M.\ Atakishiyev and M.K.\ Atakishiyeva, Theor.\ Math.\ Phys.\ 129
(2001) 1325.

\bibitem{sixdeniers} J.-M.\ Sixdeniers, K.A.\ Penson and A.I.\ Solomon, J.\ Phys.\ A 32
(1999) 7543.

\bibitem{cq01} C.\ Quesne, Ann.\ Phys.\ (N.Y.) 293 (2001) 147.

\bibitem{cq02} C.\ Quesne, J.\ Phys.\ A 35 (2002) 9213.

\bibitem{klauder01} J.R.\ Klauder, K.A.\ Penson and J.-M.\ Sixdeniers, Phys.\ Rev.\ A 64
(2001) 013817.

\bibitem{mandel} L.\ Mandel and E.\ Wolf, Optical Coherence and Quantum Optics,
Cambridge University Press, Cambridge, 1995.

\bibitem{walls} D.F.\ Walls, Nature (London) 306 (1983) 141.

\bibitem{yuen} H.P.\ Yuen, Phys.\ Lett.\ A 56 (1976) 105.

\bibitem{aragone} C.\ Aragone, G.\ Guerri, S.\ Salam\'o and J.L.\ Tani, J.\ Phys.\ A 7
(1974) L149.

\end {thebibliography}
\end{document}